\documentclass[twocolumn]{aastex6}
\pdfoutput=1 
\usepackage{amsmath,amstext}
\usepackage[T1]{fontenc}
\usepackage{apjfonts} 
\usepackage[figure,figure*]{hypcap}
\usepackage{affils}

\def\M{M$_\odot$}
\def\ergscma{\ergs\,cm$^{-2}$\,\AA$^{-1}$}
\def\kms{km\,s$^{-1}$}
\def\cmg{cm$^{2}$\,g$^{-1}$}
\def\ergs{erg\,s$^{-1}$}
\def\xlan{$X_{\rm lan}$}
\def\Mej{$M_{\rm ej}$}
\def\vk{$v_{\rm k}$}

\shorttitle{Optical spectra of the first LIGO/Virgo neutron star merger}
\shortauthors{Nicholl et al.}

\begin{document}

\title{The Electromagnetic Counterpart of the Binary Neutron Star Merger LIGO/VIRGO GW170817. III. Optical and UV Spectra of a Blue Kilonova From Fast Polar Ejecta}

\DeclareAffil{cfa}{Harvard-Smithsonian Center for Astrophysics, 60 Garden Street, Cambridge, Massachusetts 02138, USA; \href{mailto:matt.nicholl@cfa.harvard.edu}{matt.nicholl@cfa.harvard.edu}\vspace{-0.1cm}}
\DeclareAffil{columbia}{Columbia Astrophysics Laboratory, Columbia University, New York, NY 10027, USA\vspace{-0.1cm}}
\DeclareAffil{berkeley}{Departments of Physics and Astronomy, 366 LeConte Hall, University of California, Berkeley, CA 94720, USA\vspace{-0.1cm}}
\DeclareAffil{lbnl}{Nuclear Science Division, Lawrence Berkeley National Laboratory, Berkeley, CA 94720, USA\vspace{-0.1cm}}
\DeclareAffil{soar}{Southern Astrophysical Research Telescope, Casilla 603, La Serena, Chile\vspace{-0.1cm}}
\DeclareAffil{ctio}{Cerro Tololo Inter-American Observatory, National Optical Astronomy Observatory, Casilla 603, La Serena, Chile\vspace{-0.1cm}}
\DeclareAffil{ohio}{Astrophysical Institute, Department of Physics and Astronomy, 251B Clippinger Lab, Ohio University, Athens, OH 45701, USA\vspace{-0.1cm}}
\DeclareAffil{northwestern}{Center for Interdisciplinary Exploration and Research in Astrophysics and Department of Physics and Astronomy, Northwestern University, Evanston, IL 60208\vspace{-0.1cm}}
\DeclareAffil{unc}{Department of Physics and Astronomy, University of North Carolina, Chapel Hill, NC, 27599, USA\vspace{-0.1cm}}
\DeclareAffil{michigan}{Center for Data Intensive and Time Domain Astronomy, Department of Physics and Astronomy, Michigan State University, East Lansing, MI 48824, USA\vspace{-0.1cm}}
\DeclareAffil{riogrande}{Universidade Federal do Rio Grande do Sul, Instituto de F{\'{i}}sica, Porto Alegre, Rio Grande do Sul, 91501-900, Brazil\vspace{-0.1cm}}
\DeclareAffil{lowell}{Lowell Observatory, 1400 West Mars Hill Road, Flagstaff, AZ 86001, USA\vspace{-0.1cm}}
\DeclareAffil{rome}{Dipartimento di Matematica e Fisica, Universit{\`{a}} Roma Tre, via della Vasca Navale 84, 00146, Roma, Italy\vspace{-0.1cm}}
\DeclareAffil{laplata1}{Facultad de Ciencias Astron{\'{o}}micas y Geof{\'{i}}sicas, Universidad Nacional de La Plata, Paseo del Bosque, B1900FWA, La Plata, Argentina\vspace{-0.1cm}}
\DeclareAffil{laplata2}{Instituto de Astrof{\'{i}}sica de La Plata, CONICET-UNLP, CCT La Plata, Paseo del Bosque, B1900FWA, La Plata, Argentina\vspace{-0.1cm}}
\DeclareAffil{torino}{Dipartimento di Fisica, Universit{\`{a}} degli Studi di Torino, via Pietro Giuria 1, 10125, Torino, Italia\vspace{-0.1cm}}
\DeclareAffil{nucleare}{Istituto Nazionale di Fisica Nucleare, Sezione di Torino, via Pietro Giuria 1, 10125, Torino, Italia\vspace{-0.1cm}}
\DeclareAffil{bologna}{INAF-Istituto di Astrofisica Spaziale e Fisica Cosmica di Bologna, via Gobetti 101, 40129, Bologna, Italia\vspace{-0.1cm}}
\DeclareAffil{inaf-torino}{INAF-Osservatorio Astrofisico di Torino, via Osservatorio 20, I-10025 Pino Torinese, Italy\vspace{-0.1cm}}
\DeclareAffil{warwick}{Department of Physics, University of Warwick, Coventry, CV4 7AL, UK\vspace{-0.1cm}}
\DeclareAffil{kavli}{Kavli Institute for Cosmological Physics, University of Chicago, Chicago, IL 60637, USA\vspace{-0.1cm}}
\DeclareAffil{chi-astro}{Department of Physics, University of Chicago, Chicago, IL 60637, USA\vspace{-0.1cm}}
\DeclareAffil{chi-phys}{Department of Astronomy and Astrophysics, University of Chicago, Chicago, IL 60637, USA\vspace{-0.1cm}}
\DeclareAffil{penn}{Department of Physics and Astronomy, University of Pennsylvania, Philadelphia, PA 19104, USA\vspace{-0.1cm}}
\DeclareAffil{fermilab}{Fermi National Accelerator Laboratory, P.O. Box 500, Batavia, IL 60510, USA\vspace{-0.1cm}}
\DeclareAffil{nsf}{NSF GRFP Fellow\vspace{-0.1cm}}
\DeclareAffil{enrico}{Enrico Fermi Institute, University of Chicago, Chicago, IL 60637, USA\vspace{-0.1cm}}
\DeclareAffil{hubble}{Hubble Fellow\vspace{-0.1cm}}
\DeclareAffil{syracuse}{Physics Department, Syracuse University, Syracuse, NY 13244, USA\vspace{-0.1cm}}
\DeclareAffil{stsci}{Space Telescope Science Institute, 3700 San Martin Drive, Baltimore, MD 21218, USA\vspace{-0.1cm}}
\DeclareAffil{brandeis}{Department of Physics, Brandeis University, Waltham, MA 02454, USA\vspace{-0.1cm}}
\DeclareAffil{hopkins}{Department of Physics and Astronomy, The Johns Hopkins University, 3400 North Charles Street, Baltimore, MD 21218, USA\vspace{-0.1cm}}

\affilauthorlist{
M.~Nicholl\affils{cfa},
E.~Berger\affils{cfa},
D.~Kasen\affils{berkeley,lbnl},
B.~D.~Metzger\affils{columbia},
J.~Elias\affils{soar},
C.~Brice{\~{n}}o\affils{ctio},
K.~D.~Alexander\affils{cfa},
P.~K.~Blanchard\affils{cfa},
R.~Chornock\affils{ohio},
P.~S.~Cowperthwaite\affils{cfa},
T.~Eftekhari\affils{cfa},
W.~Fong\affils{northwestern},
R.~Margutti\affils{northwestern},
V.~A.~Villar\affils{cfa},
P.~K.~G.~Williams\affils{cfa},
W.~Brown\affils{cfa},
J.~Annis\affils{fermilab},
A.~Bahramian\affils{michigan},
D.~Brout\affils{penn},
D.~A.~Brown\affils{syracuse},
H.-Y.~Chen\affils{chi-astro},
J.~C.~Clemens\affils{unc},
E.~Dennihy\affils{unc},
B.~Dunlap\affils{unc},
D.~E.~Holz\affils{kavli,chi-astro,chi-phys,enrico},
E.~Marchesini\affils{laplata1,laplata2,torino,nucleare,bologna},
F.~Massaro\affils{torino,nucleare,inaf-torino},
N.~Moskovitz\affils{lowell},
I.~Pelisoli\affils{riogrande,warwick},
A.~Rest\affils{stsci,hopkins},
F.~Ricci\affils{rome,cfa},
M.~Sako\affils{penn},
M.~Soares-Santos\affils{fermilab,brandeis},
J.~Strader\affils{michigan}
}

\begin{abstract}

We present optical and ultraviolet spectra of the first electromagnetic counterpart to a gravitational wave (GW) source, the binary neutron star merger GW170817. Spectra were obtained nightly between 1.5 and 9.5 days post-merger, using the SOAR and Magellan telescopes; the UV spectrum was obtained with the \textit{Hubble Space Telescope} at 5.5 days. Our data reveal a rapidly-fading blue component ($T\approx5500$\,K at 1.5 days) that quickly reddens; spectra later than $\gtrsim 4.5$\,days peak beyond the optical regime. The spectra are mostly featureless, although we identify a possible weak emission line at $\sim 7900$\,\AA\ at $t\lesssim4.5$ days. The colours, rapid evolution and featureless spectrum are consistent with a ``blue''  kilonova from polar ejecta comprised mainly of light $r$-process nuclei with atomic mass number $A\lesssim140$. This indicates a sight-line within $\theta_{\rm obs}\lesssim45^{\circ}$ of the orbital axis. Comparison to models suggests $\sim0.03$\,\M\ of blue ejecta, with a velocity of $\sim0.3c$. The required lanthanide fraction is $\sim 10^{-4}$, but this drops to $<10^{-5}$ in the outermost ejecta. The large velocities point to a dynamical origin, rather than a disk wind, for this blue component, suggesting that both binary constituents are neutron stars (as opposed to a binary consisting of a neutron star and a black hole). For dynamical ejecta, the high mass favors a small neutron star radius of $\lesssim12$\,km. This mass also supports the idea that neutron star mergers are a major contributor to $r$-process nucleosynthesis.

\end{abstract}


\section{Introduction}

The early years of gravitational wave (GW) astronomy with the Advanced Laser Interferometer Gravitational-Wave Observatory (LIGO) have witnessed great successes in detecting the mergers of binary black holes \citep{abbott2016a,abbott2016c,abbott2017}.
However, binary neutron star (BNS) mergers, if detected, are a much more promising avenue for electromagnetic (EM) follow-up, as they are expected to produce EM signals over a wide range of frequencies and timescales \citep[see review by][]{metzger2012}. BNS mergers have long been argued to be the progenitors of short gamma-ray bursts (SGRBs) based on both theoretical viability \citep{eichler1989,narayan1992} and observations of their X-ray, optical and radio afterglows \citep{berger2014,berger2005,fox2005,hjorth2005,soderberg2006,fong2013,fong2015}. 

BNS mergers are also thought to be a promising astrophysical site for rapid neutron-capture ($r$-process) nucleosynthesis \citep{lattimer1974,eichler1989}, or even the dominant site \citep{freiburghaus1999}. This suggests an additional source of transient EM radiation is possible: an optical/IR `kilonova', powered by radioactive decays of $r$-process nuclei synthesized in the merger ejecta \citep{davies1994,li1998,rosswog1999,metzger2010}. A kilonova was likely detected in near-IR imaging following the short GRB\,130603B \citep{berger2013,tanvir2013}. Jets in core-collapse supernovae may be an alternative (or additional) site of the $r$-process \citep{winteler2012,nishimura2015}.

\begin{figure*}
\centering
\includegraphics[width=17.cm]{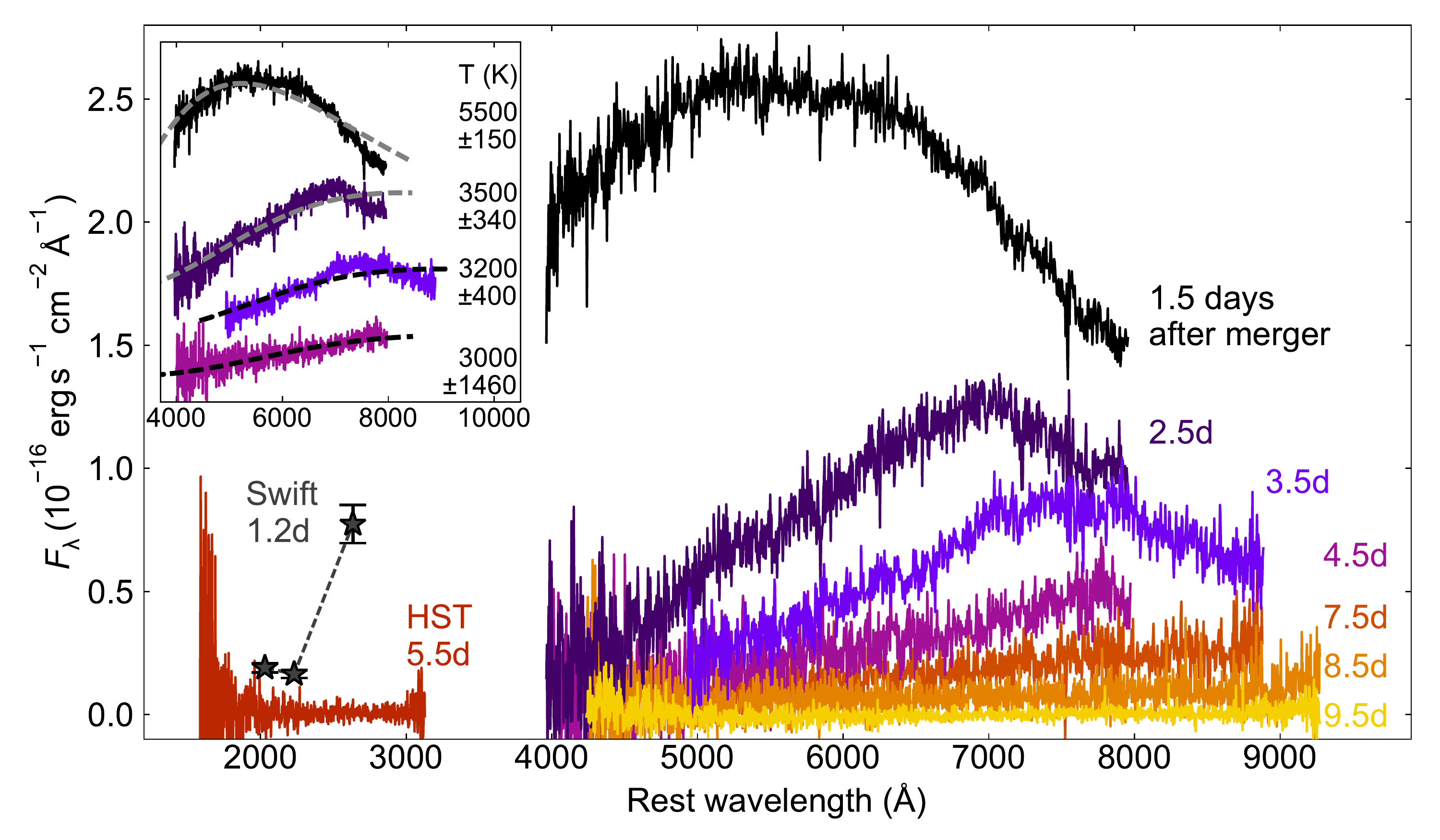}
\caption{Optical spectra of the BNS merger event GW170817. SOAR and Magellan spectra have been binned by a factor 2 for clarity. The spectra at times $\lesssim 4.5$\,d exhibit a clear optical peak that rapidly moves red. After this time, the flux is dominated by an IR component discussed in \citet{DECamPaper4}. The UV data from \textit{HST} (S/N\,$<1$, essentially an upper limit) and \textit{Swift} show blanketing at short wavelengths. Inset: blackbody fits. The early spectra are more sharply peaked than blackbody emission, due to the deficit of blue flux. At later times, the optical data are consistent with the blue tail of a $\sim 3000$\,K blackbody peaking in the near-IR.}
\label{fig:spec}
\end{figure*}

The luminosity, timescale, and spectra of a kilonova are sensitive to the opacity of the ejecta, and hence to the details of the $r$-process nucleosynthesis (e.g., the electron fraction, $Y_e$) \citep{lattimer1976,symbalisty1982}.  In particular, even a small fraction of lanthanides or actinides ($f$-shell elements) in the ejecta can increase the opacity by orders of magnitude, significantly reducing the peak luminosity and shifting the emission primarily to the infrared \citep{kasen2013,tanaka2013}. 

On 2017 August 17.53 UT, Advanced LIGO/Virgo made the first detection of gravitational waves from a neutron star binary merger, GW170817 \citep{ALVgcn,ALVdetection}, with a simultaneous SGRB detected by {\it Fermi} and \textit{INTEGRAL} (GRB\,170817A; \citealt{GBMgcn1,INTEGRALgcn}; see \citealt{DECamPaper7} for a comparison to other SGRBs). At 0.5 days after the GW signal, an EM counterpart---the first for any GW source---was identified within the $\sim 30$\,deg$^2$ localization region. The source, first announced by \citet{SWOPEgcn,SWOPEpaper}, was independently discovered by our group, using the Dark Energy Camera on the 4-m Blanco Telescope \citep{DECAMgcn,DECamPaper1}, and by several other groups \citep{DLT40gcn,DLT40paper,LCOGTgcn,LCOGTpaper,VISTAgcn,MASTERgcn}. The counterpart---variously named SSS17a, DLT17ck, and AT\,2017gfo---resides in the galaxy NGC\,4993\footnote{Throughout this paper we assume a distance to NGC\,4993 of 39.5\,Mpc and redshift $z=0.00973$ as listed in the NASA Extragalactic Database} \citep[see][for host galaxy analysis]{DECamPaper8}.

Here we report optical and UV spectra of GW170817, spanning 1.5 to 9.5 days, with dense time sampling. We demonstrate that the initial spectra are dominated by a rapidly-fading blue component that quickly evolves to the red. These properties suggest that the optical counterpart is a kilonova---the first to be observed spectroscopically. The relatively blue colour is possible for lanthanide-poor $r$-process ejecta \citep{metzger2010,barnes2013,metzger2014}, indicative of a viewing angle of $\theta_{\rm obs}\lesssim 45^\circ$ of face-on. We investigate the constraints on the mass and velocity of the polar ejecta, its composition, and the neutron star equation of state. 

\section{Observations}

Following the discovery of the optical counterpart of GW170817, we began spectroscopic observations with the 4.1-m Southern Astrophysical Research (SOAR) Telescope on 2017 August 18.97 UT. We used the Goodman High Throughput Spectrograph \citep{clemens2004} with the 400 lines-per-mm grating and 1'' slit ($R\sim 930$), beginning with the blue (`M1'; 4000--8000\,\AA) setting but switching to the red (`M2'; 5000--9000\,\AA) setup as the spectrum evolved to redder wavelengths. The slit was aligned parallel to the axis from the transient to the host nucleus, to facilitate removal of host flux. This also enabled spectroscopic analysis of the host galaxy, presented in \citet{DECamPaper8}. An atmospheric dispersion corrector was used to mitigate differential slit losses, since observations were conducted at high airmass. In total we obtained five epochs of spectroscopy with SOAR spanning $1.5-7.5$ days.

Subsequently, we  obtained optical spectra with the 6.5-m Magellan/Baade telescope equipped with the Inamori Magellan Areal Camera and Spectrograph \citep[IMACS;][]{dressler2011}. We used the 300 lines-per-mm grism ($R\sim 1100$) with a $17^\circ$ tilt for the broadest wavelength coverage, $\sim 4000-9000$ \AA). These spectra were obtained at the parallactic angle. We obtained two epochs at 8.5 and 9.5 days before the transient became too faint and unobservable due to its proximity to the Sun.

\begin{table*}
\centering
\caption{Log of optical and UV spectra \label{tab:spec}}
\begin{tabular}{cccccccccc}
\hline
MJD    & Phase$^a$  & Telescope &  Instrument  & Camera &	Grism or 	&  Exposure	& Average  &  Wavelength  & Resolution\\
    &   &    & &   &	grating	&  time (s)	& airmass  &  range (\AA)  & (\AA)\\
\hline			    		                    
57984.0 & 1.5  &  SOAR  & GHTS & Blue      &	400-M1 &   $3\times1200$ 	& 1.6 & 4000--8000     &  6  \\
57985.0 & 2.5  &  SOAR  & GHTS  & Blue      &	400-M1 &   $3\times900$ 	& 1.6 & 4000--8000     &  6  \\
57986.0 & 3.5  &  SOAR  & GHTS  & Blue      &	400-M2 &   $3\times900$ 	& 1.6 &  5000--9000    &  6  \\
57987.0 & 4.5  &  SOAR  & GHTS  & Red      &	400-M1 &   $3\times900$ 	& 1.6 &  4000--8000    &  6  \\
57988.1 & 5.5  &  \textit{HST}  & STIS  & NUV/MAMA  &  G230L	&   2000    &  ---  & 1600--3200  &  3  \\
57990.0 & 7.5  &  SOAR  & GHTS  & Blue      &	400-M2 &   $3\times900$ 	& 1.9 & 5000--9000     &  6  \\
57991.0 & 8.5  &  Magellan Baade  & IMACS  & f2  &  G300-17.5	&  $2\times1200$  & 2.0 & 4300--9300   & 6 \\
57992.0 & 9.5  &  Magellan Baade  & IMACS  & f2   &  G300-17.5	&  $2\times1350$  & 2.1 & 4300--9300   & 6  \\

\hline
\end{tabular}

$^a$ Phase in rest-frame days relative to GW signal.

\end{table*}

We processed and analyzed all spectra using standard procedures in \texttt{iraf}, including bias and flat-field corrections and background subtraction. The light profile of the host was found to be smooth at the location of the transient, and was well fit by a low-order polynomial. Wavelength calibration was performed by comparison lamp spectra, while flux calibration was achieved using standard star observations on each night. The final calibrations were scaled to match DECam photometry observed at the same time \citep{DECamPaper2}. The spectra were corrected for a Milky Way extinction $E(B-V)=0.1053$, using the dust maps of \citet{schlafly2011}, and cosmological redshift. We assume that extinction in NGC\,4993 is negligible, based on modelling by \citet{DECamPaper8}.

We additionally obtained one epoch of UV spectroscopy through Director's Discretionary Time with the \textit{Hubble Space Telescope} using the Space Telescope Imaging Spectrograph (STIS) with the NUV/MAMA detector and broad G230L grating, covering $\sim 1500$--3000\,\AA\footnote{Program GO/DD 15382, P.I.~Nicholl}. Acquisition imaging was carried out using the clear CCD50 filter. The transient is detected clearly in a pair of 90\,s CCD50 exposures. However, no trace is visible in the UV spectrum, indicating that the source is extremely UV-faint. In an effort to use all available data, we extracted the flux from the reduced 2D spectrum in an aperture centered on the source position in the acquisition image, using a recent STIS spectrum obtained with the same setup \citep{blanchard2017} to define the shape of the spectral trace. We find a 3$\sigma$ upper limit on the flux of $F_{\rm UV}\lesssim 1.5 \times 10^{-18}$\,\ergscma\ at 2500\,\AA.


\section{Spectral Properties}

Our spectroscopic time series is shown in Figure~\ref{fig:spec}. The optical spectra are unprecedented in appearance, and evolve rapidly. The earliest spectrum, at 1.5 days after the merger, shows a peak at around 5000\,\AA\ with $\lambda L_\lambda \approx 2 \times 10^{41}$\,\ergs, and a steep decline towards both UV and redder wavelengths. Photometry from the \textit{Swift} UV Optical Telescope (UVOT) shows that this steep drop continues to $\sim 2000$\,\AA\ \citep{SWIFTgcn1,SWIFTgcn2}. DECam photometry and near-IR spectroscopy presented by \citet{DECamPaper2} and \citet{DECamPaper4} show an upturn in flux above $\gtrsim 9000$\,\AA; see those works for detailed discussion of the near-IR.

The most striking aspect of these data is the rapid evolution exhibited over the first few days. On day 2.5, the optical peak has shifted redward to 7000\,\AA, while the flux drops steeply each night. The spectrum at day 4.5 appears to still show an optical peak at $\sim 7800$\,\AA\ after rebinning, but between 4.5 and 7.5 days the peak shifts completely out of the optical regime and the spectrum appears featureless.  The \textit{HST} observation, on day 5.5, shows that the early UV flux seen by \textit{Swift} has almost completely disappeared. The spectra on days 7.5--9.5 show no change in shape but continue to fade significantly from one night to the next.

These properties are not consistent with a GRB afterglow. Assuming the standard synchrotron model \citep{sari1998,granot2002}, SGRB afterglows consist of three broken power laws that generally give a much bluer optical spectrum than we observe here, and do not exhibit rapid colour evolution. The lack of a significant afterglow contribution at these epochs is supported by X-ray \citep{DECamPaper5} and radio \citep{DECamPaper6} data that indicate an off-axis jet. Models that fit the delayed onset of X-ray and radio emission imply a negligible optical contribution ($\gtrsim 30$\,mag) in the first two weeks \citep{DECamPaper5}. Apart from a SGRB afterglow, only one other channel is predicted to produce a bright optical-infrared signal in a BNS merger \citep{metzger2012}: the radioactive decay of newly synthesized $r$-process elements in the merger ejecta: i.e., a kilonova.

Next, we investigate whether the spectral energy distribution (SED) can be fit with blackbody radiation. These fits are shown in Figure \ref{fig:spec} (inset). The day 1.5 spectrum can be approximately modeled as a blackbody, with a best-fitting temperature of $\sim 5500$\,K. Taking this at face value, the luminosity requires a radius of $\sim 7\times 10^{14}$\,cm and hence an expansion velocity $\sim 0.2c$. This is consistent with the findings by \citet{DECamPaper2}, and is close to the escape velocity from neutron stars, matching theoretical predictions for the dynamical polar ejecta from the collision interface in a BNS merger \citep{hotokezaka2013,bauswein2013,sekiguchi2016}. However, we note that these are not precise values as the blackbody does not fully capture the shape of the spectrum.

On days 2.5--3.5, the blackbody fits are poor. The peak of the observed spectrum is much sharper than a blackbody of appropriate colour temperature ($\lesssim 3500$\,K). It is therefore clear that additional physics is required beyond a simple cooling of the ejecta. One possibility is that we are seeing heavy line absorption on the blue side, up to $\approx 7000$\,\AA. This requires a significantly greater line opacity than any previously observed optical transients. From day 4.5 onwards, the data can be fit with a $\approx 3000$\,K blackbody of declining flux, but as the spectral peak shifts out of the optical regime this becomes poorly constrained. The uncertainty on the temperature at 4.5 days is $\sim 50\%$, so we do not show blackbody fits to the later data.

We are not aware of any other model predictions or observed transients that match the very red colour in our spectra at $\gtrsim 2.5$ days, but kilonovae can do this because of the high optical opacity in $r$-process ejecta \citep[e.g][]{kasen2013,tanaka2013}. Therefore our preferred interpretation of our data is that they represent the first spectra obtained of a kilonova. This supports the light curve modelling in \citet{DECamPaper2} and the evidence for $r$-process material in the infrared spectrum \citep{DECamPaper4}.

\begin{figure}
\centering
 \includegraphics[width=\columnwidth]{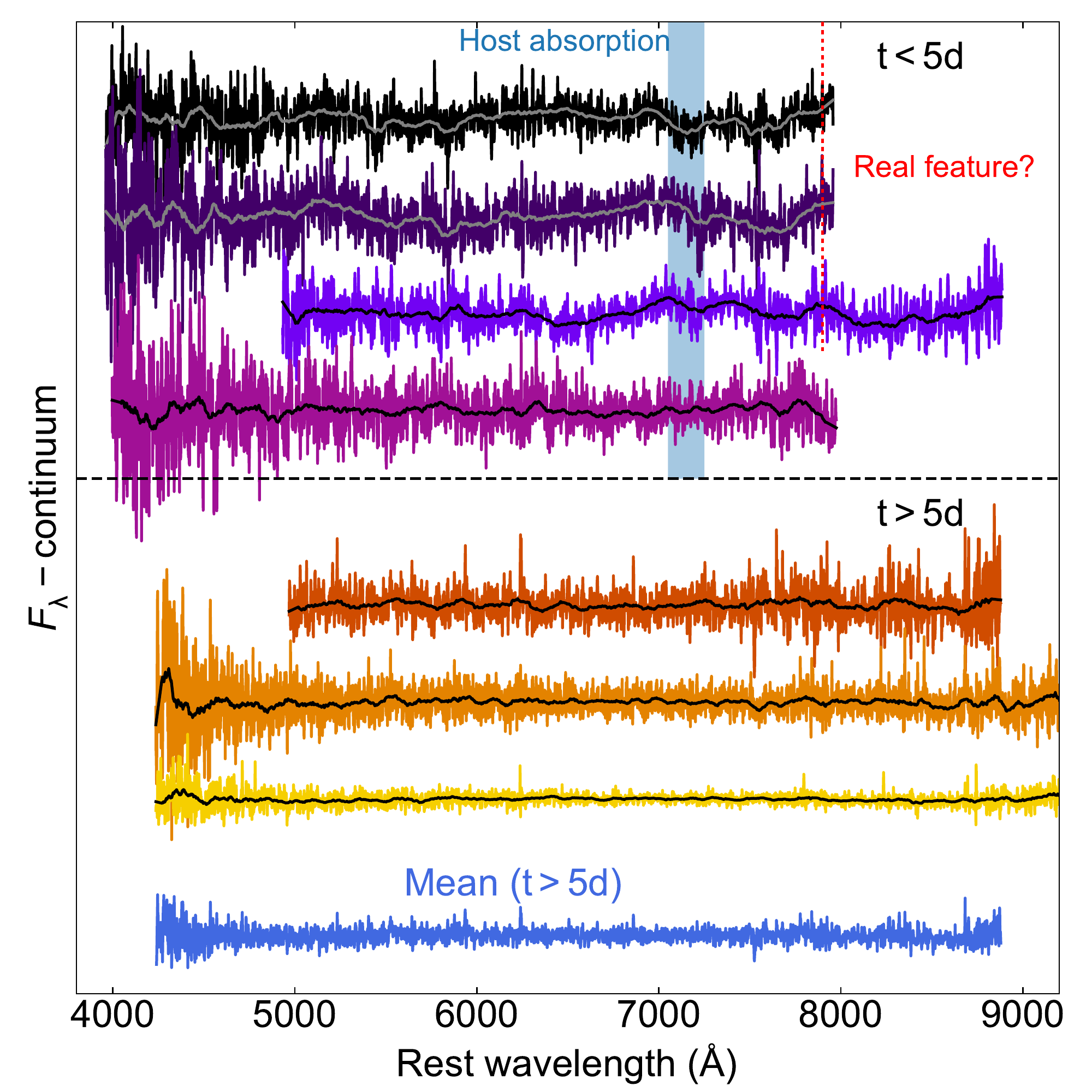}
\caption{Spectra after subtraction of a third-order polynomial fit to the pseudocontinuum. Smoothing with a Savitsky-Golay filter was applied to search for possible spectral lines. The spectra are remarkably featureless apart from a possible emission feature at $\sim 7900$\,\AA\ (unfortunately at the edge of our first two spectra). No lines are apparent at $\gtrsim 4$ days.  Similarly, no nebular features are evident up to 9.5 days after merger.}
\label{fig:sub}
\end{figure}

One remarkable aspect of our spectra is the lack of strong absorption or emission features. We search for weak spectral lines by fitting each spectra with a low-order polynomial (we experimented with third--fifth degree polynomials, finding no significant differences) and subtracting out this pseudocontinuum. We smooth each subtracted spectrum using a Savitsky-Golay filter. The results are shown in Figure \ref{fig:sub}. 

We divide the spectra into those with ($< 5$\,days) and without ($> 5$\,days) an optical peak. In both cases, any `wiggles' in the subtracted spectra are weak in comparison to the noise. Three possible features are apparent at the $\gtrsim 1\sigma$ level in the early spectra. We identify the feature at $\sim 7100$\,\AA\ with a residual from a strong host galaxy absorption trough. The spectrum at 2.5 days also exhibits a possible feature at $\sim 5100$\,\AA; this does not show up in the other spectra, and aligns with another absorption in the host galaxy spectrum. However, the first three spectra do show a possible feature at 7900\,\AA\ that we have been unable to associate with any host galaxy or reduction artefacts. Unfortunately, this is right on the edge of the wavelength range covered by our spectra on days 1.5 and 2.5, but appears to be a fairly broad emission line in our day 3.5 spectrum. We make no attempt to associate these possible features with known atomic transitions, given the modest signal-to-noise ratio and limited atomic data available for the putative $r$-process composition. 
From day 4 onwards, the spectra are apparently featureless. We search for emission lines by stacking these data, but the combined spectrum shows no significant lines. We therefore rule out nebular features brighter than $F_\lambda\approx 10^{-17}$\,\ergscma, or $L_\lambda \approx 2 \times 10^{35}$\,\ergs\,\AA$^{-1}$ ($3\sigma$).

\begin{figure}
\centering
 \includegraphics[width=\columnwidth]{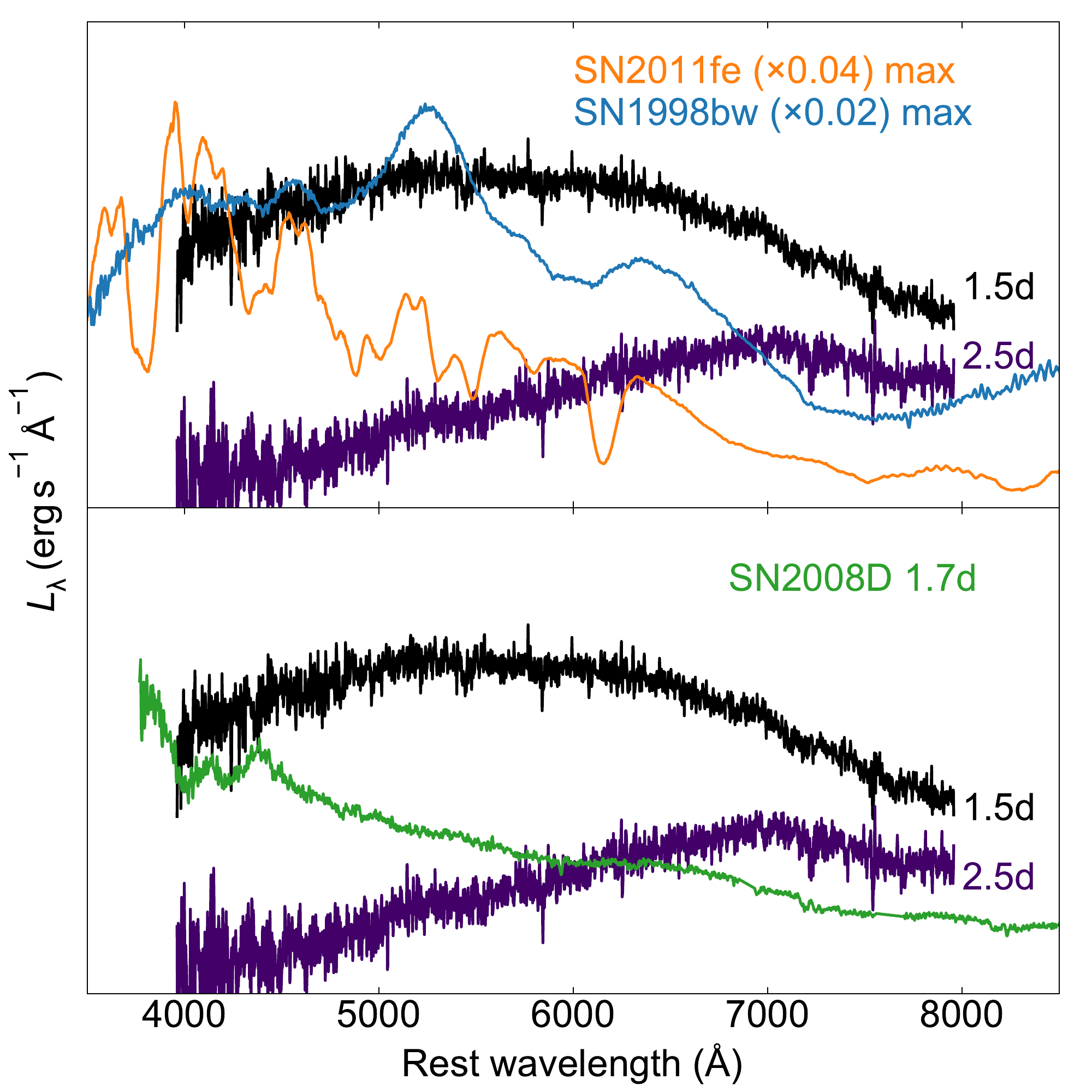}
\caption{Comparison of our optical spectra of GW170817 to supernova spectra. Top: supernovae at maximum light. The supernova opacity is dominated by iron group elements, with a velocity of $\sim 10^4$\,\kms. The GW170817 spectra do not show obvious resolved line features like the supernovae. Bottom: supernova shortly after explosion. Despite the comparable age of the ejecta from the BNS merger, the spectra are much redder than a supernova. Together, this suggests: (i) a composition with many more blended lines at blue/UV wavelengths, and (ii) significantly faster expansion velocities.}
\label{fig:ia}
\end{figure}

The lack of strong spectroscopic features, coupled with the flux suppression in the blue, suggests that the spectra are dominated by doppler-broadened blends of many overlapping atomic transitions. This is consistent with the inferred expansion velocity ($\sim 0.2c$) and the large number of blue and UV atomic lines expected for $r$-process ejecta \citep{kasen2013}. This can be seen most strikingly by comparing the spectrum to supernovae, as shown in Figure \ref{fig:ia}.

A Type Ia SN at maximum light \citep[represented by SN\,2011fe;][]{parrent2012} has an opacity dominated by iron group elements, and expansion velocities $\sim 10^4$\,\kms, lower than our estimated ejecta velocity by a factor $\sim 5$. This results in a much bluer spectrum with clearly resolved line features. A broad-lined Type Ic SN \citep[SN\,1998bw;][]{patat2001} can have expansion velocities up to $\sim 0.1c$; however, with a composition of iron group and intermediate mass elements, it is still much bluer than our observations. Finally, we compare to SN\,2008D (a Type Ib SN) at 1.7 days after explosion \citep{modjaz2009}. At a comparable phase to our spectra, the difference in colour is remarkable, with the supernova being much bluer.

\begin{figure*}
\centering
 \includegraphics[width=\textwidth]{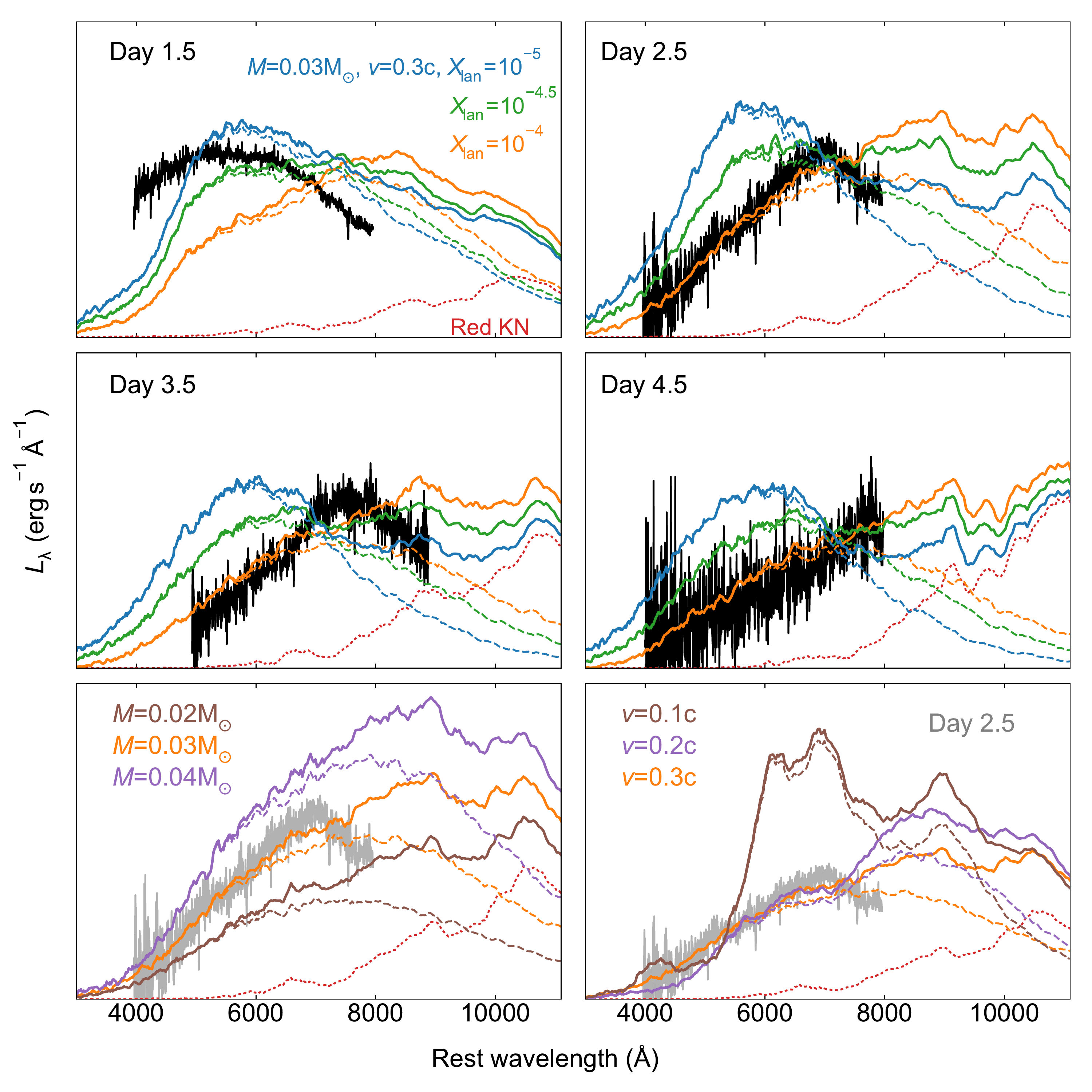}
\caption{Comparison of our optical spectra of GW170817 to kilonova models \citep{kasen2017}. Top 4 panels: Dashed lines show `blue' kilonova models for a range of lanthanide fractions ($10^{-5} < X_{\rm lan} < 10^{-4}$). The dotted line is the best-fitting red kilonova ($X_{\rm lan} = 10^{-2}$) from \citet{DECamPaper4}. Solid lines are the sum of blue and red components. An ejecta mass of \Mej\,$\approx 0.03$\,\M\ with an average velocity \vk\,$\approx 0.3c$ qualitatively reproduces the luminosity and line blending in the spectra. The rapid movement of the spectral peak from blue to red is suggestive of a concentration gradient, with a larger lanthanide fraction at lower velocity coordinate. Bottom 2 panels: The effect of varying mass and velocity away from our fiducial parameters at a fixed $X_{\rm lan}=10^{-4}$ (orange lines), compared to spectrum at day 2.5. Velocities $\gtrsim0.2c$ are required to match the lack of spectral features.}
\label{fig:mod}
\end{figure*}

\section{Comparison to Kilonova Models}
\label{sec:comp}

The significant line blanketing, mildly relativistic expansion, peak flux and rapid fading shown in the previous section are all consistent with theoretical predications for a kilonova. The initial peak in the optical and rapid decline on a $\sim$day timescale indicate that we are observing a ``blue'' kilonova, thought to occur if a significant fraction of ejected material has $Y_e\gtrsim 0.3$, suppressing the formation of lanthanides \citep{metzger2014}. This populates the first two $r$-process peaks, with mass number $A\lesssim 140$ \citep{b2fh}. In contrast, material with $Y_e\lesssim 0.2$ robustly produces lanthanides. Because of the much higher opacity, the emission from lanthanide-rich matter peaks in the infrared on a $\sim$ week-long timescale---this phenomenon is known as a ``red'' kilonova. \citet{DECamPaper4} discuss the evidence for a separate red kilonova in GW170817, while \citet{DECamPaper2} model the combined light curves of the blue and red components. Here we compare the spectrum of the optical blue component to predictions from kilonova models.

\citet{kasen2013} give approximate analytic formulae for the duration and effective temperature of the kilonova emission, in terms of ejecta mass, velocity and opacity. Using the rough temperature (5000\,K) and velocity ($0.2c$) from the blackbody fit at 1.5 days, and assuming the lifetime of the blue kilonova is $\lesssim 4-5$ days based on the lack of a distinct optical peak beyond this, we use their equation 3 to find a mass $M_{\rm ej} \sim {\rm few} \times 10^{-2}$\,\M\ in the blue ejecta. This is also consistent with the mass we derive from modelling the light curve \citep{DECamPaper2}. The corresponding (grey) opacity from equation 2 in \citet{kasen2013} is $\kappa_{\rm grey} \sim {\rm few}$\,\cmg---this is an order of magnitude greater than typical supernova opacities, but significantly less than the lanthanide opacities found by \citet{kasen2013} ($\sim 10$\,\cmg). 

Using the estimated mass and velocity as a guide, we compare to kilonova models for the spectrum, calculated using the radiative transfer code \texttt{Sedona} \citep{kasen2006}.  These models are an updated version of those from \citet{barnes2013}, and are described in detail by \citet{kasen2017}. The underlying ejecta are assumed to be spherically symmetric with uniform abundances, and in local thermodynamic equilibrium, with a density that follows a broken power-law (inner $\rho \propto v^{-1}$; outer $\rho \propto v^{-10}$). The free parameters are ejecta mass $M_{\rm ej}$, a scale velocity $v_{\rm k} = \sqrt{2E_{\rm k}/M_{\rm ej}}$, where $E_{\rm k}$ is the kinetic energy, and the lanthanide fraction \xlan.

We compare the kilonova models to our spectra in Figure \ref{fig:mod}. These models are not fit to the data in a formal sense; we simply choose the closest representative models from the grid calculated by \citet{kasen2017}. In each case, we show a model for the blue ejecta, as well as a separate red component that matches the observed near-IR spectrum \citep[see][]{DECamPaper4}. The red kilonova model has an ejecta mass 0.035\,\M\ and a velocity 0.1$c$, with a lanthanide fraction \xlan\,$=10^{-2}$. The red emission falls far below the observed flux at wavelengths $\lesssim 8000$\,\AA, so in kilonova models the optical luminosity must be dominated by the blue component.

None of the models provide a very close fit to the data. In particular, we did not find any model that could produce such a sharp peak as we see in the data on day 2.5. However, by examining qualitative similarities and the effects of varying each parameter, we can still gain useful insights.

To match the lack of features in the optical data, models need a high velocity in the blue component---those with velocities $\lesssim 0.1c$ show numerous resolved lines \citep{kasen2015,kasen2017}. The best-fitting models here have an average velocity \vk\,$=0.3c$. 
A larger mass of $r$-process material provides a greater luminosity. Matching our spectra requires a blue ejecta mass \Mej\,$\approx 0.03$\,\M, in good agreement with our simple analytic estimate.

We note that there is some degeneracy between \Mej\ and \vk\ in matching the spectral properties, which should be resolved in future using a finer grid spacing \citep{kasen2017}, but models where the parameters deviate by more than $\sim 50\%$ from our fiducial values seem to provide a poor fit to the data. Reassuringly, the parameters we find from the spectra are in good agreement with the light curve fits in \citet{DECamPaper2}. These estimates for mass and velocity can be tested: for the parameters given here, \citet{DECamPaper6} predict a bright radio signal $\sim 5$ years post-merger.

The model spectra are very sensitive to \xlan, which largely sets the peak wavelength through its impact on the opacity. The first spectrum at 1.5 days, with a peak at $\sim 5000$\,\AA, requires \xlan\,$\lesssim10^{-5}$. In fact, the data are bluer than even the most lanthanide-poor model shown at 1.5 days, perhaps indicating that additional physics is required at early times. At this epoch, the near-IR luminosities of the blue models are comparable to that of the red model, but the blue kilonova fades much faster and makes only a minor contribution above $\gtrsim 10000$\,\AA\ on the following nights. This is in agreement with the findings of \citet{DECamPaper4}.

At later epochs, the model with \xlan\,$=10^{-5}$ gives a poor fit, peaking too far to the blue. The spectrum on day 2.5 is better modelled with \xlan\,$\approx 10^{-4.5}-10^{-4}$. By day 3.5 (and subsequently; spectra beyond day 4.5 not shown), the spectrum is most consistent with \xlan\,$=10^{-4}$. Overall, the spectroscopic evolution is indicative of a gradient in lanthanide abundance, with a modest fraction \xlan\,$\sim 10^{-4}$ in the inner ejecta and a much lower fraction at the surface. 

This comparison serves as a first step to estimate the approximate mass, velocity and lanthanide fraction of the ejecta, but there remain significant discrepancies between data and model. Much more detailed modelling with further fine tuning will be required to derive better constraints on the key properties. For example, models that deviate from spherical symmetry and local thermodynamic equilibrium should be explored.

\section{Implications}\label{sec:diss}

First and foremost, the apparent discovery of a kilonova associated with GW170817, and modelling of its physical properties, supports the theory that BNS mergers are an important site of the $r$-process \citep[see also][]{DECamPaper2,DECamPaper4}. The optical emission is consistent with ejecta consisting primarily of species with\ $A\lesssim 140$, i.e.~the first two $r$-process abundance peaks. \citet{DECamPaper4} argue that a near-IR component also seen in GW170817 includes heavier $r$-process elements. Assuming that our kilonova interpretation is correct, we now examine the consequences of observing this blue component.

Lanthanide-poor ejecta can form in two ways: either through the shock-heating of material at the point of contact, ejected with high velocity and entropy \citep{oechslin2007,bauswein2013,hotokezaka2013}; or through late-time escape via an accretion disk wind, if the merger product (a hyper- or supramassive neutron star) avoids collapse to a black hole long enough to provide a sustained neutrino flux that raises the electron fraction to $Y_e\gtrsim 0.3$ \citep[the required lifetime is $\gtrsim 100$\,ms;][]{metzger2014,kasen2015}.

In both cases, this material is expected only within $\lesssim 45^\circ$ of the orbital axis of the binary \citep[e.g.][]{wanajo2014,goriely2015,sekiguchi2016,foucart2016,radice2016}. In the orbital plane,  material is instead ejected by tidal forces and is expected to have $Y_e\lesssim 0.2$. Hence the tidal material will be rich in lanthanides, producing only the near-IR red kilonova.

Even if a significant fraction of the mass is ejected in the polar direction with a high $Y_e$, seeing the blue emission may be strongly dependent on the viewing angle: at angles $\theta_{\rm obs}\gtrsim 45^\circ$, the higher opacity of the lanthanide-rich ejecta can obscure the blue ejecta from view. The previous best kilonova candidate, GRB\,130603B, appeared to be a manifestation of the more isotropic red kilonova \citep{berger2013,tanvir2013}\footnote{However, the presence of blue kilonova emission is challenging to constrain through GRB follow-up observations, because it is generally dimmer than the optical afterglow for on-axis observers \citep{metzger2012}}.
Modelling of the X-ray and radio data for GW170817 by \citet{DECamPaper5} and \citet{DECamPaper6} suggests a viewing angle $\theta_{\rm obs}\gtrsim 20^\circ$, so combining this with our restriction from the blue emission gives a tight constraint $20^\circ \lesssim \theta_{\rm obs}\lesssim 45^\circ$. This can help to overcome the distance-inclination degeneracy in the gravitational wave signal, and derive a more precise distance (and hence cosmological parameters; \citealt{H0paper}). However, the polar ejecta may be visible for a wider range of angles if it expands faster than the equatorial tidal ejecta.


We have found evidence for high velocities ($0.2-0.3c$). This has important implications for the origin of the high-$Y_{e}$ matter that can give rise to blue kilonova emission. {While winds from the accretion disk around the post-merger remnant can eject the requisite amount of mass (e.g.~\citealt{fernandez2013,metzger2014,perego2014,siegel2017}), the predicted wind velocities are too low ($<0.1c$).  Our estimated outflow velocity is closer to that expected for the dynamical polar ejecta from the shocked interface at the point of collision (e.g.~\citealt{oechslin2007}), which would support a dynamical origin for the blue emission.} \citet{DECamPaper2} also find high velocities for the blue ejecta, and reach a similar interpretation. In the case of a merger between a black hole (BH) and a neutron star, the only source of high-$Y_e$ ejecta is a disk wind, as there is no contact interface \citep[e.g][]{fernandez2017,kasen2017}. Therefore the presence of a significant mass of fast blue ejecta would disfavour an interpretation of GW170817 in which one of the binary members is a black hole. This is a crucial advantage of EM follow-up, as GW observations alone cannot distinguish between NS-NS binaries and NS-BH binaries.

If the ejecta are indeed dynamical as we infer, the total ejected mass is most sensitive to the neutron star radius: the more compact the neutron star, the closer the binary members can approach each other, and hence the higher the orbital velocity at merger, leading to stronger shocks that heat and eject more material \citep{hotokezaka2013,bauswein2013}. 
This is an important point, since the radius of a neutron star is one of its most challenging properties to measure, with the current uncertainty range from $\sim11$ to 14 km \citep[see review by][]{ozel2016}.

Simulations show that $\sim 10^{-2}$\,\M\ can be ejected for small neutron star radii $\lesssim 11$\,km, whereas the ejecta mass is an order of magnitude lower for larger radii  $\gtrsim 13$\,km. Our interpretation therefore favours a small neutron star radius (i.e.~a soft equation of state), though more modelling will be required, both on the mass ejection from compact object mergers and on reproducing the properties of our spectra, to confirm if the picture we suggest is the unique solution.

\section{Conclusions}

We have presented a high-cadence time series of optical spectra for the first EM counterpart of a LIGO/Virgo GW source, the BNS merger GW170817. Our observations reveal an optical excess fading over a timescale of a few days. The earliest spectrum, 1.5 days after the merger, peaks at 5000\,\AA\ with a luminosity $\lambda L_\lambda \approx 2 \times 10^{41}$\,\ergs. On subsequent nights, the peak of the spectrum moves quickly to redder wavelengths as the luminosity declines over a timescale $\lesssim 5$ days. The spectra are largely featureless, while the spectral slope, and our UV spectrum from \textit{HST}, indicates severe line blanketing.

The luminosity and colour temperature, as well as the implied velocity, are qualitatively consistent with models for blue kilonovae---optical transients powered by the radioactive decay of $r$-process material ejected from the merger. This adds support to the long-standing suspicion that BNS mergers are the main site of $r$-process nucleosynthesis. The optical luminosity requires a low fraction of lanthanide-series elements that would otherwise suppress the optical flux through line opacity. Such lanthanide-poor ejecta are expected to be visible only within $\theta_{\rm obs}\lesssim 45^\circ$ of the orbital axis (with lanthanide-rich ejecta outside of this solid angle), therefore constraining our viewing angle to the system.

Spectral models need a high velocity $v\sim 0.3c$ to match the featureless spectra through line blending, suggesting that the blue material was ejected dynamically rather than in a disk wind. Polar dynamical ejecta are only predicted for NS-NS binaries, not for NS-BH binaries, so our interpretation favours the former. The mass required to match the luminosity in the spectrum is \Mej\,$\approx 0.03$\,\M, similar to that derived from the light curve \citep{DECamPaper2}, with \xlan\,$\approx10^{-4}$ and fewer lanthanides at higher velocity coordinate. As simulations suggest that this mass is close to the maximum amount of dynamical polar ejecta that can result from a BNS merger \citep{oechslin2007,bauswein2013,hotokezaka2013}, this could point to a compact neutron star and therefore a soft equation of state.

In future observing runs, LIGO/Virgo should detect many more BNS mergers. The discovery of a blue/optical transient associated with GW170817, in addition to the expected IR emission, suggests that optical follow-up searches may have an easier time finding the counterparts than previously thought \citep[e.g.][]{cowperthwaite2015}. Spectral analysis of future targets will be essential in determining the prevalence and luminosity functions of blue kilonovae \citep[see][]{DECamPaper7}, allowing us to map out the mass and velocity distributions of the ejecta. This in turn will greatly increase our understanding of the mass ejection and nucleosynthesis, placing much firmer constraints on the constituent neutron stars.


\acknowledgments
We thank an anonymous referee for a very swift and helpful review that improved the paper. The Berger Time-Domain Group at Harvard is supported in part by the NSF through grants AST-1411763 and AST-1714498, and by NASA through grants NNX15AE50G and NNX16AC22G. Based on observations obtained at the Southern Astrophysical Research (SOAR) telescope, which is a joint project of the Minist\'{e}rio da Ci\^{e}ncia, Tecnologia, Inova\c{c}\~{a}os e Comunica\c{c}\~{a}oes (MCTIC) do Brasil, the U.S. National Optical Astronomy Observatory (NOAO), the University of North Carolina at Chapel Hill (UNC), and Michigan State University (MSU). This paper includes data gathered with the 6.5 meter Magellan Telescopes located at Las Campanas Observatory, Chile.
JS is supported by a Packard Fellowship. This research has made use of the NASA/IPAC Extragalactic Database (NED), which is operated by the Jet Propulsion Laboratory, California Institute of Technology, under contract with the National Aeronautics and Space Administration.

\bibliographystyle{yahapj}

\end{document}